A. Kozhanov et al. October 2009                                                                                                1A. Kozhanov et al. October 2009                                    1

# Resonant coupling of coplanar waveguides with ferromagnetic tubes.

A. Kozhanov[1], D. Ouellette[1], M. Rodwell[1], D. W. Lee[2], S. X. Wang[2] and S. J. Allen[1]
[1]*California Nanosystems Institute, University of California at Santa Barbara, Santa Barbara, CA, 93106*
[2]*Department of Materials Science and Engineering, Sanford University, Stanford, CA, 94305*
(Received     )

Resonant coupling of coplanar waveguides is explored by wrapping proximate shorted ends of the waveguides with micron size ferromagnetic $Co_{90}Ta_5Zr_5$ tubes. Ferromagnetic resonance and up to 7 outer surface modes are identified. Experimental results for these contorted rectangular tubes are in good agreement with micromagnetic simulations and model calculations of magnetostatic modes for an elliptical ferromagnetic tube. These results indicate that the modes are largely determined by tube topology and dimensions but less so by the detailed shape. (PACS: 76.50.+g)
Monolithic micro and nanoscale filters, delay lines and resonators are potentially important for high frequency electronics. Here we explore metallic ferromagnetic micron scale magnetostatic wave structures in the form of tubes wrapped around coplanar waveguides. Microwave devices based on magneto-static spin waves in insulating ferrimagnetic materials like yttrium-iron garnet (YIG)[1] have long been explored and developed. However, future micro and nano scale spin wave based devices may benefit from exploiting ferromagnetic metals that are more easily deposited, processed and nanofabricated than ferrimagnetic oxides. Further, ferromagnetic metals like CoTaZr, CoFe and CoFeB have nearly an order of magnitude larger saturation magnetization than typical ferrimagnets.[2] As a result, they will support higher, shape defined, zero magnetic field resonances and consequently intrinsically faster response.

Recent theoretical[3,4,5] and experimental[6,7,8] work has focused on the magnetization dynamics in ferromagnetic nanotubes. Ferromagnetic nanotubes could serve as magnetic cores in nano scale transformers as well as active elements of tunable high frequency filters. Several theoretical models predict existence of quantized surface modes of magnetostatic oscillations in the ferromagnetic nano tubes magnetized along the axis of the tube[4,5]. The experiments of Mendach et al. observed ferromagnetic resonance in a Permalloy tube, the lowest order and essentially spatially uniform mode. They did not report on the rich spectrum of standing magneto-static waves that circulate around the tube.[8].

This letter describes excitation and detection of quantized surface magnetostatic oscillations in the ferromagnetic tubes formed by wrapping metallic ferromagnetic film around exciting and detecting coupling loops at ends of co-planar waveguides. Resonances are displayed that are defined by the magnetostatic modes indexed by periodic boundary conditions around the tube despite their contorted geometry. Such a structure could form the building block for filtering elements at microwave frequencies.

The ferromagnetic tube coupler was fabricated in the following manner. 200nm thick ferromagnetic $Co_{90}Ta_5Zr_5$ films were sputtered onto Si/SiO$_2$ wafers, lithographically patterned into two xxx x xxx micron rectangles and covered with an insulating SiO$_2$ layer. (A saturation magnetization of $M_s$=1.2T and a coercive field $H_c$~2 Oe were measured on an unpatterned $Co_{90}Ta_5Zr_5$ film using a vibrating sample magnetometer.) Coupling loops formed by the shorted ends of a pair of coplanar waveguides were positioned over $Co_{90}Ta_5Zr_5$ rectangles. The structure was then covered with a 100 nm thick SiO$_2$ insulating layer and holes etched to allow a subsequent top $Co_{90}Ta_5Zr_5$ layer to complete the magnetic circuit. The top $Co_{90}Ta_5Zr_5$ layer was sputtered on the resist covered structure and lifted-off. This process resulted in two shorted coplanar waveguides wrapped together by a $Co_{90}Ta_5Zr_5$ film (Fig.1). The "tubes" thus formed are of course not circular but topologically equivalent to a cylinder and provide a closed magnetic circuit (Fig. 1b,c) that captures the magnetic fields produced by the high frequency currents flowing in the shorted ends of the coplanar waveguides. Focused ion beam etching (Fig.1c) exposes the cross section imaged in the SEM micrograph.

The closed magnetic circuit effectively couples the two shorted ends of the coplanar waveguides only when the tube magnetization is oriented along the axis of the tubes, along the shorting lines of the coplanar waveguides. Only then can the microwave magnetic fields induce changes in the magnetization and couple to the magnetostatic oscillations in the tubes.

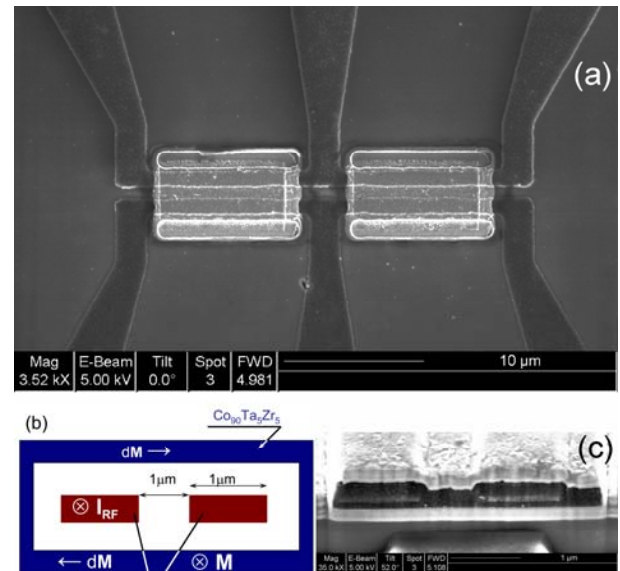

Fig. 1. Fabricated structure SEM micrograph (a), profile scheme(b), SEM micrograph of structure cross section.(c)

S-parameters were measured at room temperature using Agilent 8720ES vector network analyzer operating from 0.05 to 20 GHz. Only $S_{21}$, the ratio of high frequency voltage at terminals 2 to the input high frequency voltage at terminals 1, is analyzed in the following discussion. The test devices were positioned on the narrow gap of small electro-magnet that provided magnetic field bias up to 1000 Oe. By comparing the S-parameters at disparate bias magnetic fields, the magnetic field independent instrument response can be effectively removed to expose the S-parameters related to the magneto-static mode coupling of exciting and detecting wires. See Fig. 2.

H=0 are not observed (The "ferromagnetic" state[7] with magnetization aligned along the tube is not stable at H=0 in the particular rectangular tube used in these experiments.)

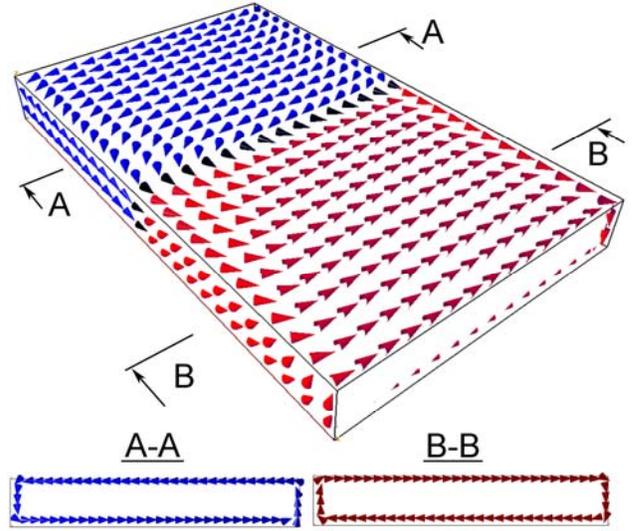

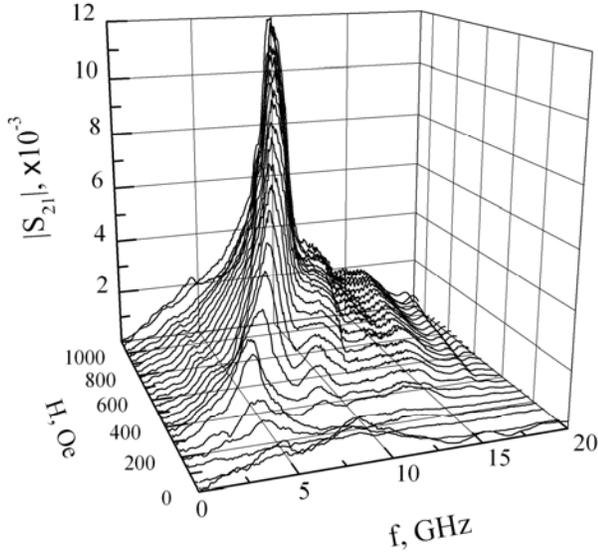

Fig. 2. Frequency and magnetic field dependence of |S21| measured with the fabricated structure.

Fig. 3. Schematic of magnetization alignment in the tube at $H_x$=0.

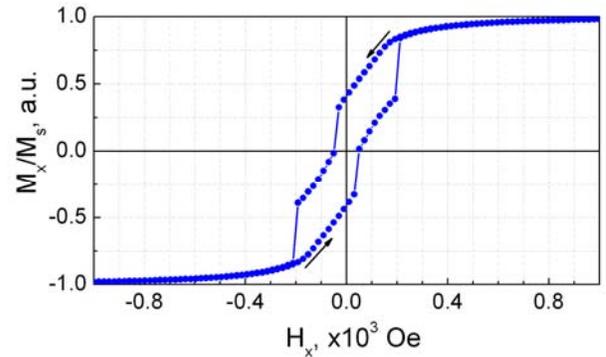

In the absence of external magnetic field, $|S_{21}(f)|$ has a number of irreproducible peaks that are strongly dependent on the history of the bias magnetic field. With increase of the external magnetic field directed along the axis of the tube ($H_x$), the magnitude of these peaks decreases and related magnetic coupling disappears at $H_x$~100 Oe.

At $H_x \geq 200$ Oe we detect a strong peak followed by a series of smaller peaks. These peaks shift towards higher frequencies and grow in magnitude with increase of $H_x$. Transmission at the lowest frequency peak reaches $|S_{21}|$~0.012 at $H_x$=988 Oe.

These results suggest that at ~ 100-200 Oe there is a change in the magnetization distribution. Micromagnetic simulations were carried out to investigate the magnetization alignment within our structures at magnetic field values |H|≤1kOe. We used a rectangular $Co_{90}Ta_5Zr_5$ tube as the model with dimensions similar to the dimensions of the fabricated ferromagnetic tubes. Micromagnetic structure was simulated by solving the Landau-Livshitz-Gilbert equation using LLG Micromagnetics Simulator.

The results of the simulations are shown on Figures 3 and 4. The ground state at $H$=0 is described by a double vortex[7] state. The magnetization is circularly oriented around the tube perimeter pointing in a clockwise direction on one end, counter-clock wise direction on the other end, with a domain wall in the center. This is similar to what was found by Lee et al. for circular ferromagnetic nanotubes[7]. For the rectangular tube some of the quasi stable states apparent in the in circular tube at

Fig. 4. Results of micromagnetic simulations: hysteresis curve of the rectangular ferromagnetic tube.

In our experiment the paired vortex state of our ferromagnetic tube will result in a complex configuration in which most of the magnetic moments are parallel to the exciting high frequency magnetic fields produced by the RF currents in the waveguide. Only the very narrow area of the domain wall between the paired vortex states magnetic moments will have some alignment along the tube axis and couple to the exciting magnetic fields. We speculate that they are responsible for the irreproducible low amplitude peaks in $|S_{21}(f)|$ at $H$=0.

The simulation indicates that with sufficient field, $H_x$, the tube is magnetized along the axis; this is indicated by the saturation in the hysteresis curve (Fig.4). This proceeds either by domain wall widening or formation of in plane magnetization vortex. The details are not important for the present discussion but following the jump of the $M(H)$ curve at $H$≈200 Oe we assume the magnetic moments are largely pointing along the tube axis. Then the high frequency magnetic field is perpendicular to the magnetic moments and can excite the various standing magnetostatic waves.

Starting from magnetic field values of ~200 Oe we detect a series of resonant peaks in $|S_{21}(f)|$ whose frequency increases with increasing $H_x$ (Fig.6). The development of the systematic structure above 200 Oe shown in Fig. 6 confirms the results of the simulation: above 200Oe, the magnetization of the rectangular tube is well aligned with the tube axis.

We identify up to 8 peaks in $|S_{21}(f)|$ and display them versus external magnetic field in Figure 6. We use model by Popov and Zavislyak[5] to analyze the experimental data. This model calculates standing magnetostatic wave modes in a ferromagnetic tube with elliptical profile magnetized along the tube axis. Two classes of solution appear: standing spin waves on the inner and on the outer surfaces of the tube. We assign the observed peaks to the 7 *outer* standing magnetostatic modes plus the lowest, and essentially uniform, ferromagnetic resonance. The frequency for the latter uniform mode of the elliptical nanotubes coincides with the frequency of the ferromagnetic resonance[8] of an infinite ferromagnetic film and described by $f \sim \gamma(H + 4\pi M)^{1/2}$. The other modes describe standing magnetostatic waves with a whole number of wavelengths within the outer perimeter of the tube. The inner modes shown with the dashed lines in the Fig.6 do not seem to appear in the experimental data. We can offer no explanation.

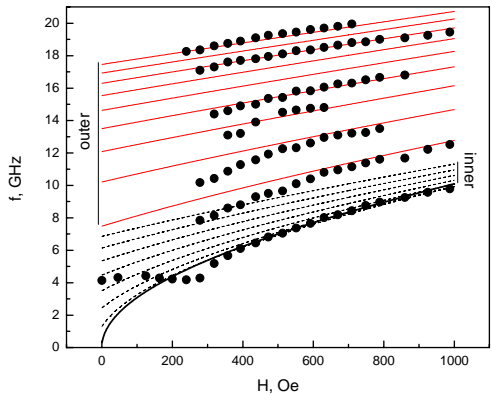

Fig. 6. $|S_{21}|$ peaks frequency dependence on the applied magnetic field: experiment (circles), theory for tubes with elliptical profile (lines)[5].

The profile of the fabricated ferromagnetic tubes is not at all elliptical but has gross distortions where the CoTaZr film goes over the coupling loops. Remarkably, we conclude that response is controlled by the topology and scale: the detailed shape is not critical.

In summary, we fabricated and measured microwave transmission through coplanar waveguides coupled by novel rectangular ferromagnetic tubes. We identified ferromagnetic resonance and up to 7 outer surface magnetostatic oscillation modes guided by models of the magnetostatic oscillations in ferromagnetic elliptical nanotubes.

These structures are potentially important as high frequency tunable filters. The frequency of the lowest and strongest peak is the ferromagnetic resonance as described earlier. The frequency of the higher modes are defined by the tube geometry (diameter, wall widths). As interesting as the surface modes are, the strong ferromagnetic mode is probably the most useful.

In order to make the filter work efficiently we will increase the input inductance of the filter either by increasing the number of exciting wires inside the magnetic tube (winding) or more conveniently by increasing the tube length while keeping other dimensions the same. Lengthening the tubes will introduce strong shape anisotropy. The ground state will have the magnetization along the tube without external bias and a strong resonance transmission at *H=0*. Resonance could be fine tuned by "on circuit board" fields or self fields produced by DC currents flowing in the wires encased by the tubes.

The authors are grateful to Andrew Cleland for the use of the vector network analyzer and hosting aspects of this work in his laboratory. This work is supported by NERC via the Nanoelectronics Research Initiative (NRI), by Intel Corp. and UC Discovery at the Western Institute of Nanoelectronics (WIN) Center.